\documentclass[10pt, conference,]{IEEEtran}
\IEEEoverridecommandlockouts
\usepackage{cite}
\usepackage{amsmath,amssymb,amsfonts}
\usepackage{booktabs}
\usepackage{algorithmicx}
\usepackage[noend]{algpseudocode}
\usepackage{graphicx}
\usepackage{textcomp}
\usepackage{xcolor} 
\usepackage{algorithm}
\usepackage{tabularx}
\usepackage{makecell}
\usepackage{easyReview}
\usepackage{multirow}
\usepackage{subcaption}
\usepackage{environ}
\usepackage{xspace}
\usepackage{hyperref}
\usepackage{titlesec}

\titlespacing{\section}{0pt}{*1}{*0.5} 
\def\BibTeX{{\rm B\kern-.05em{\sc i\kern-.025em b}\kern-.08em
    T\kern-.1667em\lower.7ex\hbox{E}\kern-.125emX}}

\ifodd 0

\else

\fi 

\NewEnviron{mycomment}{} 

\ifodd 0
\newcommand{\needrev}[1]{{\color{green}#1}}
\newcommand{\rev}[1]{{\color{blue}#1}}
\newcommand{\newrev}[1]{{\color{red}#1}}
\else
\newcommand{\needrev}[1]{#1}
\newcommand{\rev}[1]{#1}
\newcommand{\newrev}[1]{#1}
\fi 

\newcommand{\name}{KubeSpace\xspace}

\begin{document}


\title{\name: A Low-Latency and Stable \newrev{Control Plane for LEO Satellite Container Orchestration}\\
\thanks{This work was supported by the National Key R\&D Program of China under Grant No. 2023YFE0116600. The corresponding author is Wenjun Zhu.}
}

\author{\IEEEauthorblockN{Zhiyuan Zhao, Jiasheng Wu, Shaojie Su, Wenjun Zhu, Yue Gao}  \IEEEauthorblockA{Institue of Space Internet, Fudan University, China \\ College of Computer Science and Artificial Intelligence, Fudan University, China }}

\maketitle

\begin{abstract}

\newrev{Low earth orbit~(LEO) satellites play a pivotal role in global connectivity—delivering high‑speed Internet, cellular coverage, and massive IoT support. With ever‑growing onboard computing and storage resources, LEO satellites herald a new cloud paradigm: space cloud computing. While container orchestration platforms (e.g., Kubernetes) excel in terrestrial data centers, they are ill‑suited to LEO satellite networks, featuring geographic dispersion and 
\rev{frequent handovers}. Those features bring high latency and intermittent \rev{management}, leading to control plane failure in container orchestration.} 
To address this, we propose \newrev{\name,} a low-latency and stable control plane specifically designed for container orchestration on LEO satellites. \newrev{\name combines two key innovations: a distributed ground‑control‑node architecture that binds each satellite to its nearest controller for uninterrupted
\rev{management}, and an orbit‑aware placement with dynamic assignment strategy that further minimizes communication latency and handover frequency.}
Extensive experiments based on real satellite traces demonstrate that compared to existing solutions, \name reduces the average management latency of satellite nodes by 59\% without \rev{any management interruption time.}



\begin{mycomment}

\end{mycomment}

\begin{mycomment}






\end{mycomment}

\begin{mycomment}












\end{mycomment}

\begin{mycomment}

\end{mycomment}

\end{abstract}

\begin{IEEEkeywords}
LEO Satellites, Low-Latency Control Plane, Seamless Handover, Space Cloud Computing, Kubernetes
\end{IEEEkeywords}

\section{Introduction}


In recent years, low earth orbit~(LEO) satellite constellations—such as Starlink\cite{url_starlink}, Kuiper\cite{url_kuiper}, and OneWeb\cite{url_oneweb}—have developed rapidly~\cite{yuan2024satsense,peng2025sigchord,zhao2024leo,yuan2023graph}. These constellations consist of thousands of satellites equipped with high-speed inter-satellite links~(ISLs)\cite{starlink_isl} and ground-satellite links~(GSLs)\cite{starlink_gsl}, enabling network services such as high-speed Internet, cellular coverage, and large-scale IoT to be delivered globally\cite{darwish2022location,xiao2022leo,zhou2023aerospace}. Meanwhile, with increasing onboard computing and storage capabilities\cite{buckley2022radiation,sun2025intra,giuffrida2021varphi,rapuano2021fpga}, LEO satellite constellations are emerging as a computing service platform operating in space\cite{bhattacherjee2020orbit,bleier2023space,wang2023satellite1}. A variety of applications tailored for LEO satellites, such as onboard AI inference models\cite{denby2023kodan,zhang2022progress,russo2022using} and lightweight 5G core networks\cite{su2025skyoctopus,liu2024democratizing,li2022case}, are emerging. These applications are increasingly deployed and managed using container orchestration platforms such as Kubernetes\cite{bhosale2024krios,shangguang2024first,wang2023satellite2}, improving deployment flexibility and the efficiency of onboard resource management.

\begin{figure}[t]
    \centering
    \vspace{-0.25cm}
    \includegraphics[width=1\linewidth]{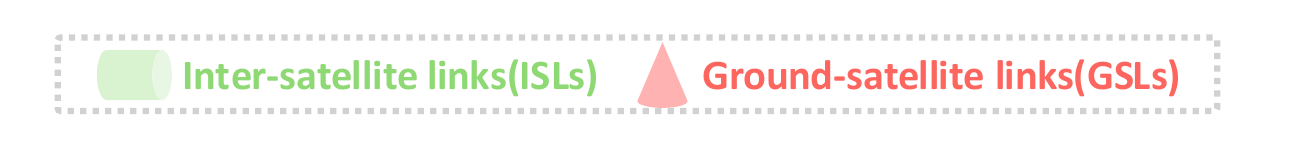}

    \begin{subfigure}[t]{0.49\linewidth}
        \centering
        \includegraphics[width=\linewidth]{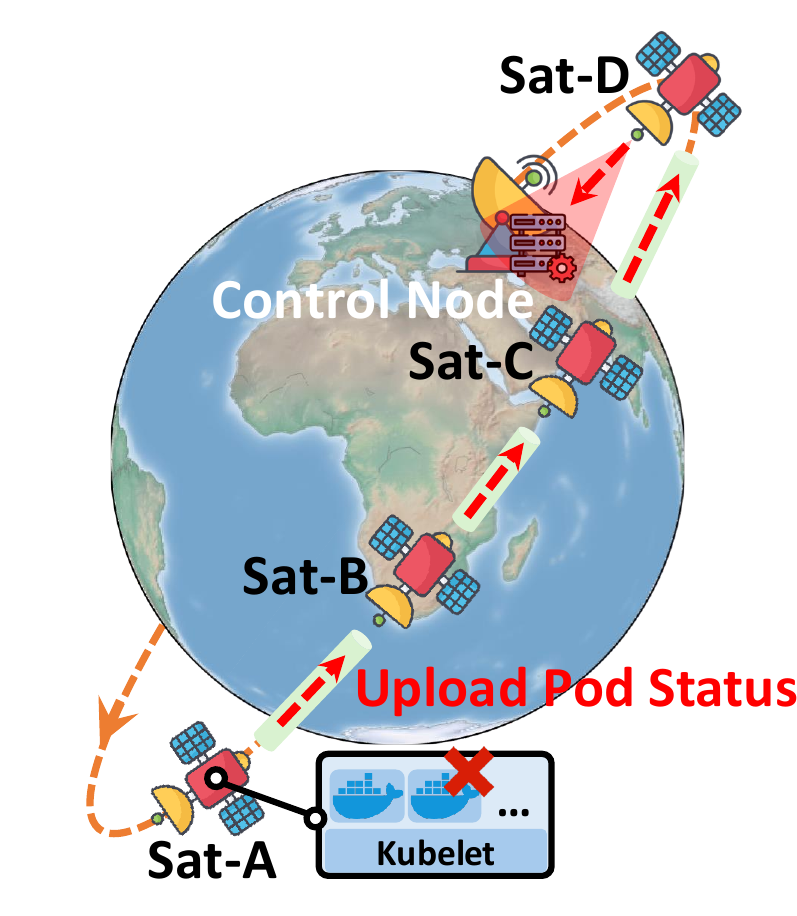}
        \caption{Single control node: high management latency.}
        \label{intro1}
    \end{subfigure}
    \hfill
    \begin{subfigure}[t]{0.49\linewidth}
        \centering
        \includegraphics[width=\linewidth]{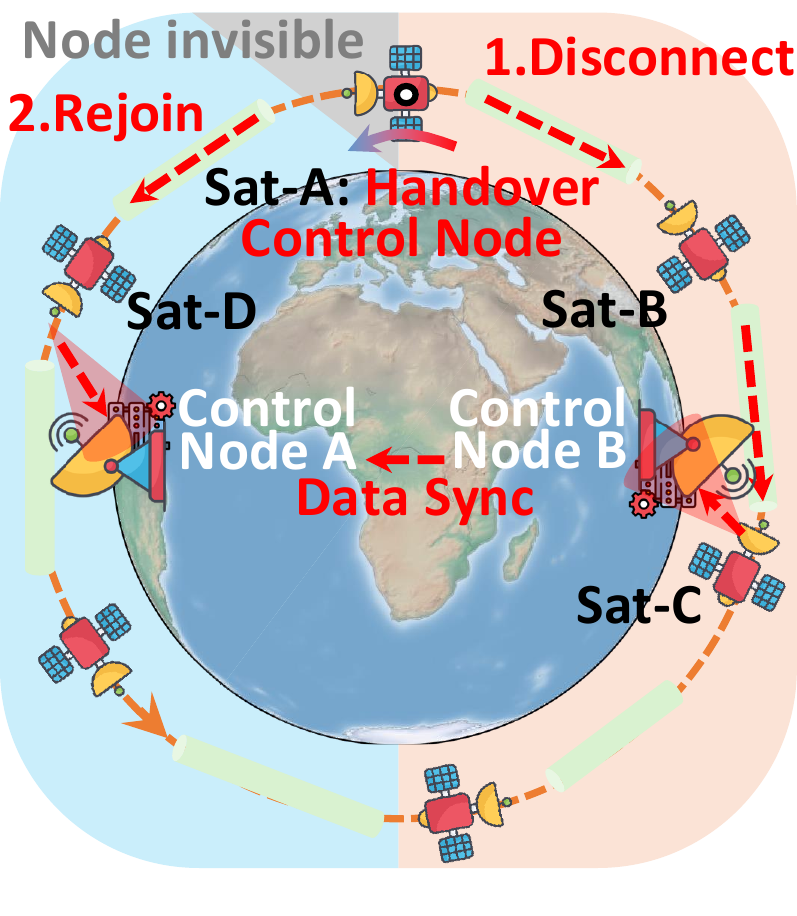}
        \caption{Multi control node: management interruption during handovers.}
        \label{intro2}
    \end{subfigure}

    \caption{Ground control node deployment for LEO satellite container orchestration: methods and challenges.}
    \vspace{0cm}
    \label{fig:intro}
\end{figure}

However, the control plane of traditional container orchestration platforms struggles to efficiently manage satellite resources and onboard containers. As illustrated in Fig.~\ref{intro1}, if a single ground station or data center serves as the sole control node, communication between certain satellites and the control node may require dozens of inter-satellite hops. This results in latencies of several hundred milliseconds, severely limiting the control plane’s ability to respond promptly to on-orbit container events.

A straightforward optimization strategy is to designate multiple ground stations as control nodes, allowing satellites to connect to the geographically closest control node. However, as shown in Fig.~\ref{intro2}, traditional multi-controller architectures do not support seamless handover of worker nodes between different control nodes. When a satellite performs a handover between control nodes, it must first leave its current cluster before joining a new one. During this process, the control plane loses visibility of the satellite, and onboard application containers are disrupted. Moreover, due to the large number and high mobility of LEO satellites, such handover events occur frequently and have become the norm. Consequently, the control plane is forced to frequently handle disruptions in satellite node management, thereby making it difficult to maintain stable control over onboard resources and containers.

This paper focuses on container orchestration platforms for LEO satellites using multiple control nodes, aiming to build a control plane with both low latency and high stability. We propose KubeSpace, which optimizes the multi control node architecture to enable seamless control node handovers for satellites, avoiding management interruptions during control transitions. Furthermore, we design a placement and dynamic allocation strategy that efficiently selects a small subset of ground stations as control nodes and leverages predictable satellite trajectories to assign each satellite an optimal controller, reducing both communication latency and handover frequency.

We implemented a KubeSpace prototype based on Kubernetes v1.31.10 and conducted extensive experiments using real LEO satellite constellation data. The experimental results show that KubeSpace achieves a 59\% reduction in satellite management latency, an 84\% decrease in control node handover time, and completely eliminates management interruptions during satellite handovers, compared to traditional container orchestration platforms.

Contributions of this paper can be summarized as follows:

\begin{itemize}

\item We are the first to reveal the challenges faced by container orchestration platforms in the context of LEO satellite deployments, where the control plane struggles to balance low latency and high stability in managing satellite nodes.

\item We propose KubeSpace, which provides a low-latency and stable control plane for container orchestration platforms by leveraging an optimized multi control node architecture and enhanced mechanisms for control node placement and allocation.

\item We implemented a prototype of KubeSpace and conducted extensive experiments, which demonstrate that it significantly reduces the latency of satellite node management and completely eliminates management interruption time during control node handovers.

\end{itemize}

The rest of this paper is organized as follows. Section II presents the background and motivation. Section III provides an overview of KubeSpace. Section IV details the design of its three key components. Section V describes the experimental setup and results. Section VI reviews related work. Section VII discusses limitations and future work. Finally, Section VIII presents the conclusion.

\section{Background and Motivation}

\begin{figure}[]
	\centering
	\includegraphics[width=1.0\linewidth]{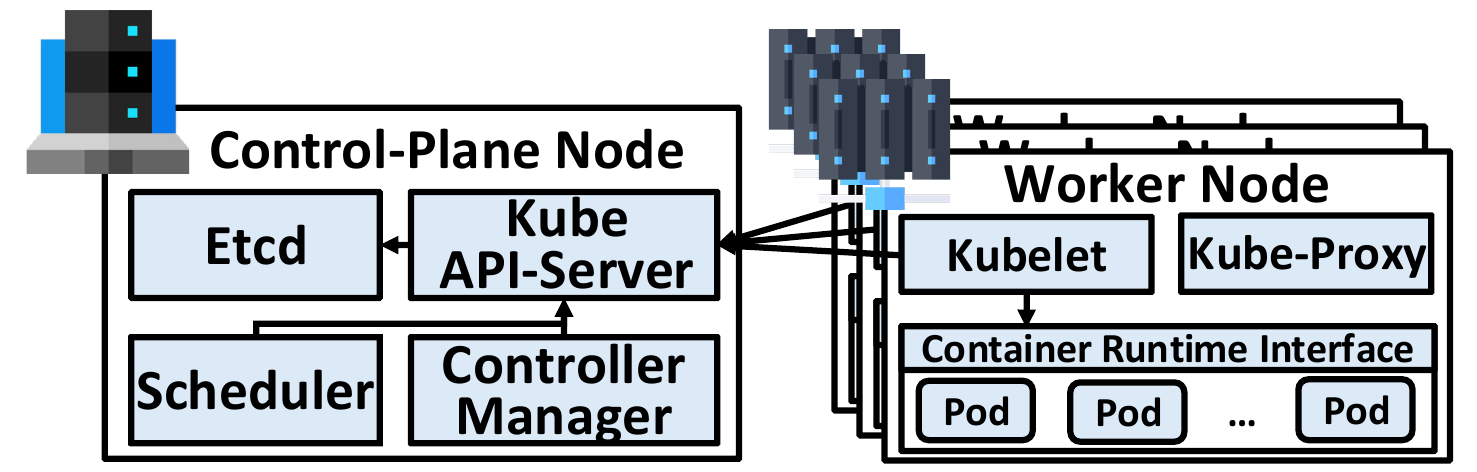}
	\caption{Architecture of Kubernetes.}
	\label{bg1}
\end{figure}

\subsection{Container Orchestration Platforms}

Container orchestration platforms originated in data centers, where Internet giants such as Google, Facebook, and Amazon used such platforms to manage tens of thousands of servers and the millions of containerized applications running on them. Today’s mainstream cluster management framework, Kubernetes\cite{kubernetes1310}, was inspired by Google’s Borg platform\cite{verma2015large}. It provides a set of control plane components and node agents to enable unified management of compute, storage, and other resources within a server cluster, and offers services such as declarative deployment and resource scheduling for containerized applications.

As shown in Fig.~\ref{bg1}, Kubernetes divides servers into two roles: control-plane nodes and worker nodes. To deploy a containerized application, the user submits a resource manifest (e.g., a YAML file) to the API Server, which stores the request data in the Etcd database. The Scheduler then selects an appropriate node based on resource constraints and scheduling policies and completes the binding process. The Controller Manager ensures that the actual state of the system converges to the desired state defined in the manifest. On each worker node, the Kubelet retrieves the Pod information assigned to it from the API Server, uses the container runtime to start the containers, and periodically reports the status of the node and its containers back to the control plane.

To ensure high availability of the control plane, data centers typically deploy multiple control nodes that collectively form a unified control plane. This architecture relies on the Etcd database to maintain data consistency. However, prior studies have shown that due to Etcd’s dependence on low-latency communication, this approach is not suitable for high-latency wide-area network environments\cite{url_etcd_hardware,jeffery2021rearchitecting,jeffery2023mutating,url_etcd_berops,url_etcd_slickfinch}. An alternative multi-controller architecture, which is the focus of this paper, divides large-scale server infrastructures into multiple clusters, each managed by an independent control node. In terrestrial network environments, mature multi-cluster solutions such as Karmada\cite{karmada2025} have been developed, enabling unified monitoring and container deployment across clusters managed by different control nodes.

\subsection{Container Orchestration in LEO Satellite Constellations}

\begin{figure}[t]
    \centering
    \vspace{-0.25cm}

    \begin{subfigure}[t]{0.69\linewidth}
        \centering
        \includegraphics[width=\linewidth]{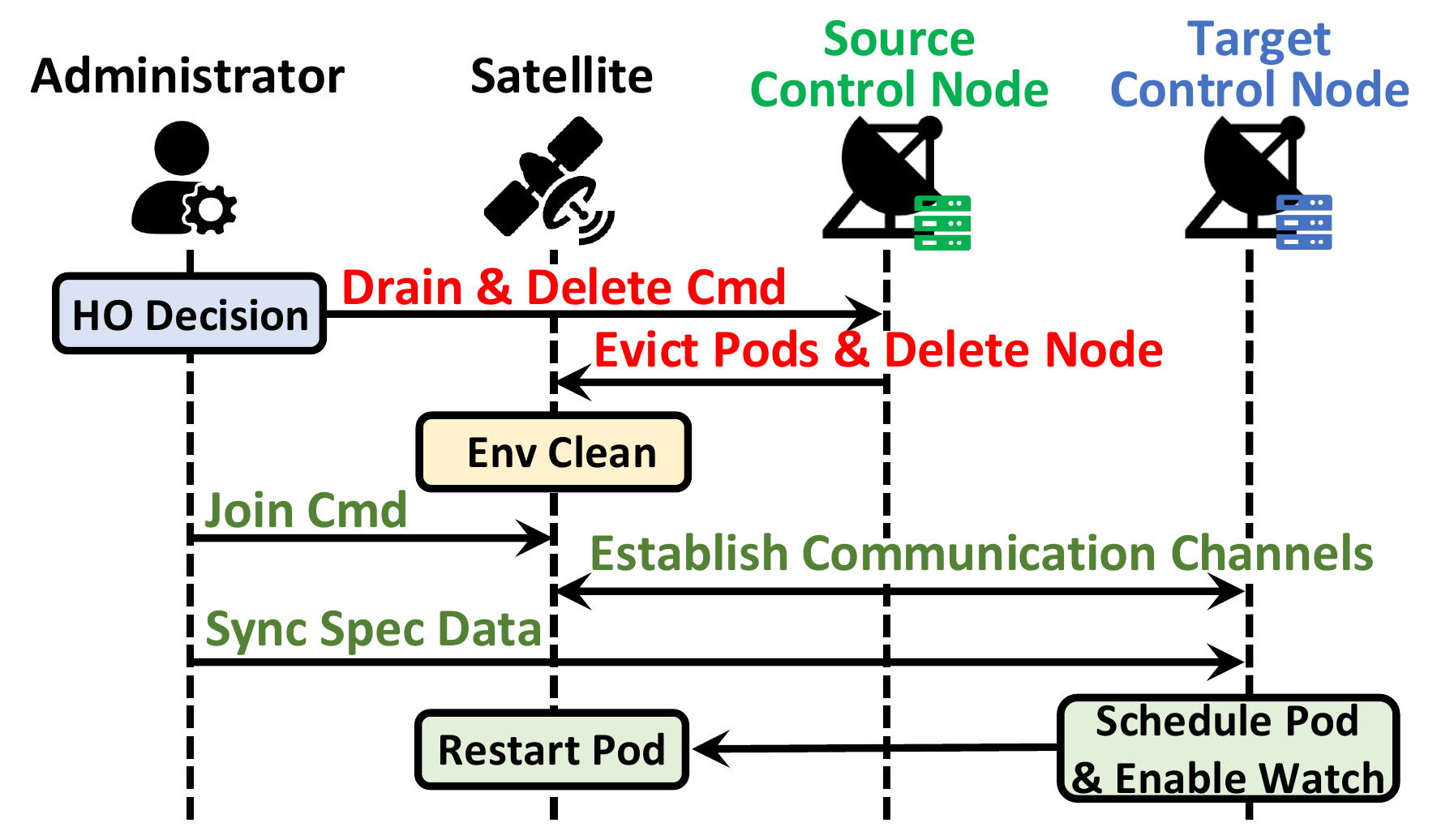}
        \caption{Workflow of control node handover.}
        \label{bg2}
    \end{subfigure}
    \hfill
    \begin{subfigure}[t]{0.28\linewidth}
        \centering
        \includegraphics[width=\linewidth]{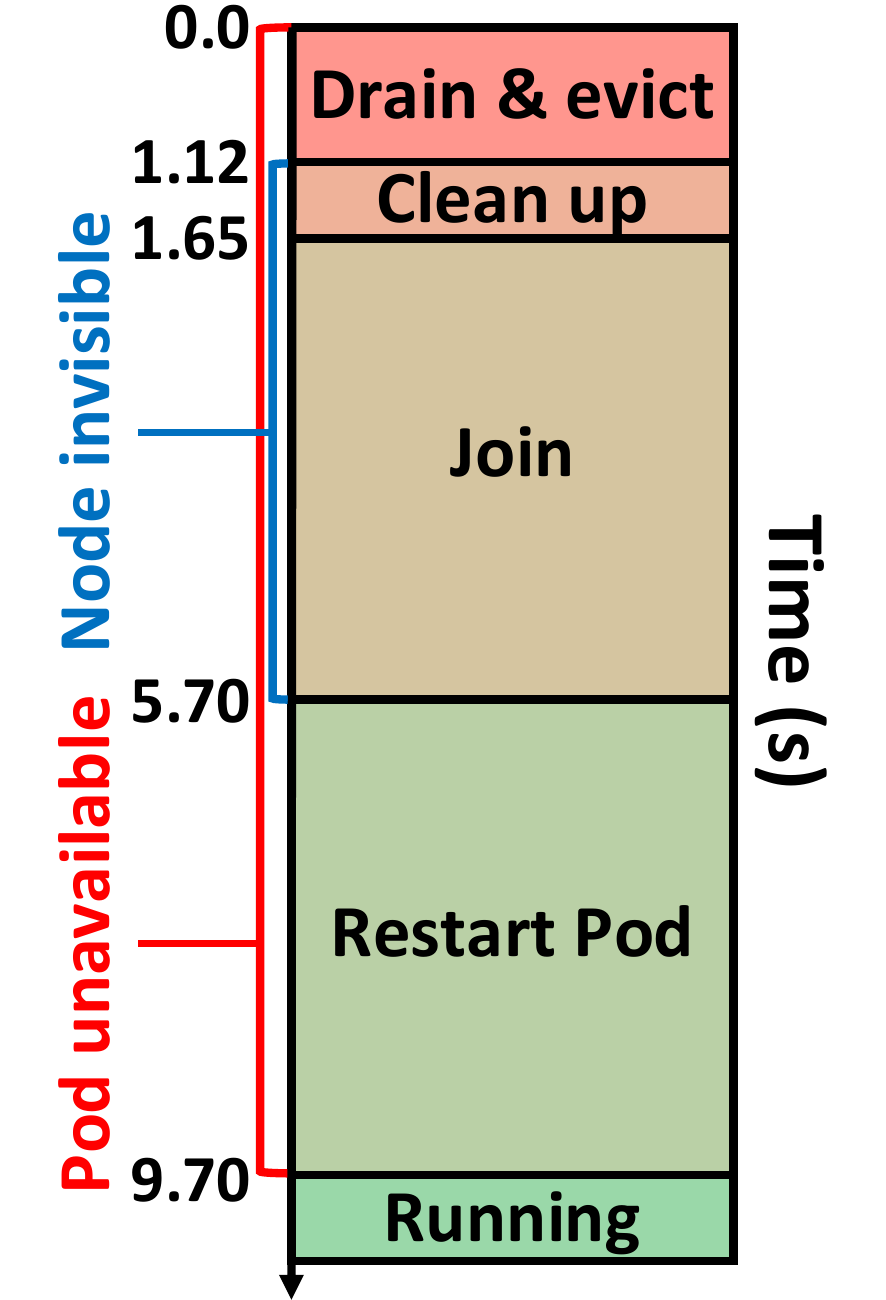}
        \caption{Time overhead.}
        \label{bg3}
    \end{subfigure}

    \caption{Handover process in traditional platform.}
    \label{fig:bg2_3}
\end{figure}

LEO satellite constellations, as a large-scale and fast-moving space-based computing infrastructure, pose significant challenges for traditional ground-based container orchestration platforms employing multiple control nodes. Although this architectural choice can help reduce management latency to some extent, existing solutions generally lack mechanisms to support the handover of worker nodes across different control nodes.

As shown in Fig.~\ref{bg2}, when a satellite needs to handover to a new control node, it must first leave the current cluster and then rejoin a new one\cite{url_k8s_drain_node,url_k8s_kubeadm_join}. First, the administrator marks the satellite as unschedulable under the current control node and evicts all Pods running on it, before removing the satellite from the cluster. The satellite then clears its local cluster configuration, performs mutual authentication with the new control node, and completes the node registration process to be brought back under the control plane's management. If any Pods are still required to run on the satellite, their scheduling metadata must be synchronized to the new control node, and the scheduling process must be re-executed.

As shown in Fig.~\ref{bg3}, assume the worker node is running only a lightweight nginx container. Even when the handover occurs within a wired local area network with a round-trip latency of less than 1 millisecond, the nginx service still requires approximately 9.7 seconds to become available again. Of this time, about 5.7 seconds are spent on stopping the container, cleaning up the environment, and completing the reconnection and re-registration with the new control node. During the transition—from disconnecting from the old control node to connecting to the new one—the node remains invisible to the control plane for approximately 4.5 seconds.

\begin{table}[t!]
  \centering
  \caption{Simulated daily control node handover overhead.}
  \vspace{0cm}
  \setlength{\tabcolsep}{0.7pt}
  \renewcommand{\arraystretch}{1.3}
    \begin{tabular}{>{\centering\arraybackslash}p{1.5cm} 
                    >{\centering\arraybackslash}p{1.3cm} 
                    >{\centering\arraybackslash}p{2cm} 
                    >{\centering\arraybackslash}p{1.7cm} 
                    >{\centering\arraybackslash}p{2.1cm}}
    \toprule
    \makecell{Constellation \\ Name} & \makecell{Total \\ Handovers} & \makecell{Avg. Duration \\ per Handover (s)} & \makecell{Total Node \\ Invisibility (h)} & \makecell{Total Pod \\ Unavailability (h)} \\
    \midrule
    Starlink & 44287 & 8.35 & 84.99 & 156.00 \\
    Kuiper   & 35782 & 8.35 & 68.67 & 126.04 \\
    Oneweb   & 15966  & 8.36 & 30.68 & 56.29 \\
    \bottomrule
  \end{tabular}
  \label{tab:addlabel}
\end{table}


Based on the constellation configurations of Starlink\cite{starlinkLEO}, Kuiper\cite{kuiperLEO}, and OneWeb\cite{onewebLEO}, along with an inter-satellite link scheme in which each satellite connects to four neighboring satellites\cite{bhosale2024krios,su2025skyoctopus}, we built a simulated constellation system. In the simulation, we deployed two ground control nodes at the equator, at longitudes 0° and 180°. The tc tool was used to simulate communication latency between satellites and ground nodes. Using this setup, we evaluated the handover overhead under a traditional multi control node architecture. As shown in Table I, using Starlink as an example, a total of 44,287 handovers occurred in one day, with each satellite experiencing an average of 28 handovers. Each handover took approximately 8.35 seconds. The total node unavailability time across all satellites over one day was 84.99 hours, and the cumulative Pod service interruption time reached 156 hours.

In conclusion, implementing container orchestration with multiple control nodes in LEO satellites is not a straightforward task. A new multi control node architecture is urgently needed to ensure that satellites connect to nearby control nodes and maintain uninterrupted management. In addition, it remains a critical challenge to determine the appropriate number and placement of ground stations for deploying control nodes, and to dynamically assign the optimal control node during satellite movement, in order to minimize communication latency and fully realize the benefits of a multi control node setup.

\section{Design Overview}

This paper presents KubeSpace, a control plane designed to meet both demands for low latency and high stability in container orchestration frameworks for LEO satellite constellations. It achieves this through two key innovations.

On the one hand, KubeSpace provides an optimized multi control node architecture with seamless control node handover for satellites. Specifically, on the control plane side, we introduce a mechanism to identify and assign satellite control ownership, and design a process for synchronizing satellite management data and transferring control authority based on Kubernetes-native Custom Resource Definitions (CRDs) and the watch mechanism. On the satellite side, we incorporate certificate preloading and hot handover of control channels. These mechanisms work in concert to effectively prevent service and management interruptions that typically occur during control node handovers in traditional platforms.

On the other hand, to further reduce communication latency between satellites and control nodes, we propose a control node placement and dynamic assignment approach. First, KubeSpace introduces a two-step selection algorithm to efficiently choose a subset of control nodes from a large number of ground stations, with the objective of minimizing the maximum satellite-to-ground communication latency across all topologies during the entire constellation operation period. The algorithm first exploits the regularity of satellite-to-ground topologies to reduce the number of topologies considered during selection via clustering. Then, a K-center algorithm optimized with local search is applied to generate a practical placement scheme within a short time. Second, the assignment of control nodes is delegated to the satellites, allowing them to independently predict control node handover times based on their TLE data\cite{tle}. Additionally, to avoid unnecessary handovers triggered by minor fluctuations in satellite-to-control-node distances, we introduce a distance threshold in the control node selection process.

\begin{figure}[t]
    \centering
    \includegraphics[width=\linewidth]{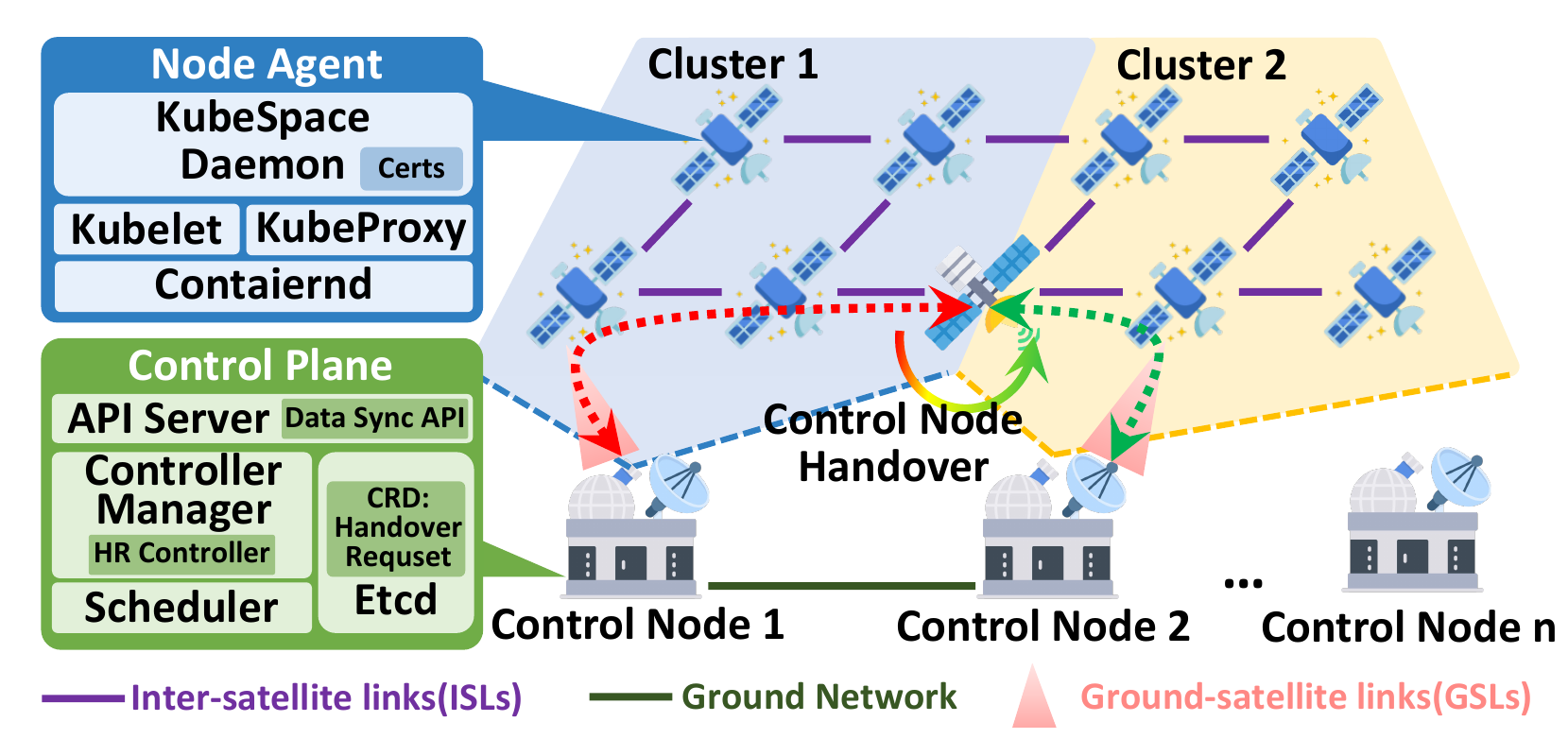}
    \caption{Multi control node architecture of KubeSpace.}
    \label{de1}
\end{figure}

\section{Design of KubeSpace}

In this section, we provide a detailed introduction to KubeSpace. We first present its multi control node architecture and the design that enables seamless control node handover for satellites. Then, we introduce two mechanisms designed to further reduce management latency and handover frequency: the Control Node Placement Algorithm~(CNPA) and the Control Node Assignment Algorithm~(CNAA).

\subsection{Multi Control Node Architecture and Handover Mechanism}


As shown in Fig.~\ref{de1}, the multi control node architecture proposed by KubeSpace involves multiple ground stations, each independently deploying a full set of Kubernetes control plane components and managing its own isolated cluster. The Kubelet agents deployed onboard the satellites function similarly to those in conventional Kubernetes clusters: they periodically report status to the currently connected ground control node and receive task assignments from it. 

Based on this foundation, we propose a state-based scheme for identifying and transitioning the control relationship between satellites and control plane nodes, which consists of the following three design aspects.


\noindent\textbf{State-based node control binding and filtering.} To clearly define each control node’s authority over satellite nodes, KubeSpace introduces a new field in the node status called ControlPlaneBindingState. This field specifies the control relationship between a satellite node and a control node, and can take one of four possible states:

\begin{itemize}
  \item \textbf{Bound}: The node is fully managed by this control node.
  \item \textbf{Binding}: Control is in the process of being transferred to this control node.
  \item \textbf{Releasing}: Control is being offboarded from this control node.
  \item \textbf{Released}: The node has been fully released from this control node's management.
\end{itemize}

By default, the Kubernetes controller manager and scheduler consider all registered nodes when performing health checks and scheduling decisions. To prevent interference among control nodes, we introduce a filtering step after retrieving the node list, excluding nodes whose ControlPlaneBindingState is not Bound. This prevents the controller manager from treating remote nodes as unhealthy and avoids scheduling pods to nodes outside its authority.


\noindent\textbf{Consistency Guarantees for Control Handover.} In a standard Kubernetes cluster, a worker node must retrieve the cluster's root certificate from the API server upon first joining, in order to verify the legitimacy of the control plane’s server certificate. It then uses a bootstrap token to request the issuance of its own certificate from the control plane, thereby establishing a secure communication channel.

We extend this process in two ways. On the one hand, all control nodes share a common key pair for issuing certificates to satellite nodes. This design allows a satellite node, once authenticated by any control node, to connect to any other control node without repeating the authentication process. On the other hand, once all ground control nodes are deployed and their certificates are generated, we preload the root certificate along with the server certificates of all control nodes onto the satellite. This eliminates the need for remote certificate retrieval during cluster join or control handover, thereby reducing connection latency.

To prevent the new control node from mistakenly recognizing a satellite node as a newly joined node during control handover—potentially causing configuration resets or container restarts—we extend the API server interface, which allows control nodes to synchronize the configuration data of satellite nodes, including the states of associated pods. Additionally, we assign a unique Pod CIDR to each satellite in advance based on its index, ensuring that multiple satellites can join the cluster concurrently without IP range conflicts. Once the satellite completes identity authentication, its initial state is synchronized across all control nodes.

\noindent\textbf{Seamless Node Control Handover Process.} To enable seamless handover between control nodes, we introduce a new custom resource type named HandoverRequest, along with a controller called HRController deployed on each control node. The entire handover process is orchestrated using Kubernetes’ native watch mechanism to ensure timely coordination.

On each satellite, we deploy a dedicated daemon called KubeSpace, which maintains mappings between control nodes and their respective certificate sets and cluster configuration files. When appropriate, KubeSpace initiates a control handover by creating a HandoverRequest resource on its current control node. This triggers the HRController on that node, which then begins the handover process.

The functionality of the kubelet is logically divided into two parts: one manages container runtime operations, while the other uses the official Kubernetes Go client library to report status and establish a watch channel for receiving control updates from the API server. The latter is critical for enabling control node handover. To support runtime flexibility, we encapsulate the kubelet’s clientset as a hot-swappable pointer. Upon handover, the kubelet initializes a new clientset and replaces the active pointer, allowing it to transition to the new control node without interrupting its control communication channel.

\begin{figure}[t]
 \vspace{0cm}
    \centering
    \includegraphics[width=1.0\linewidth]{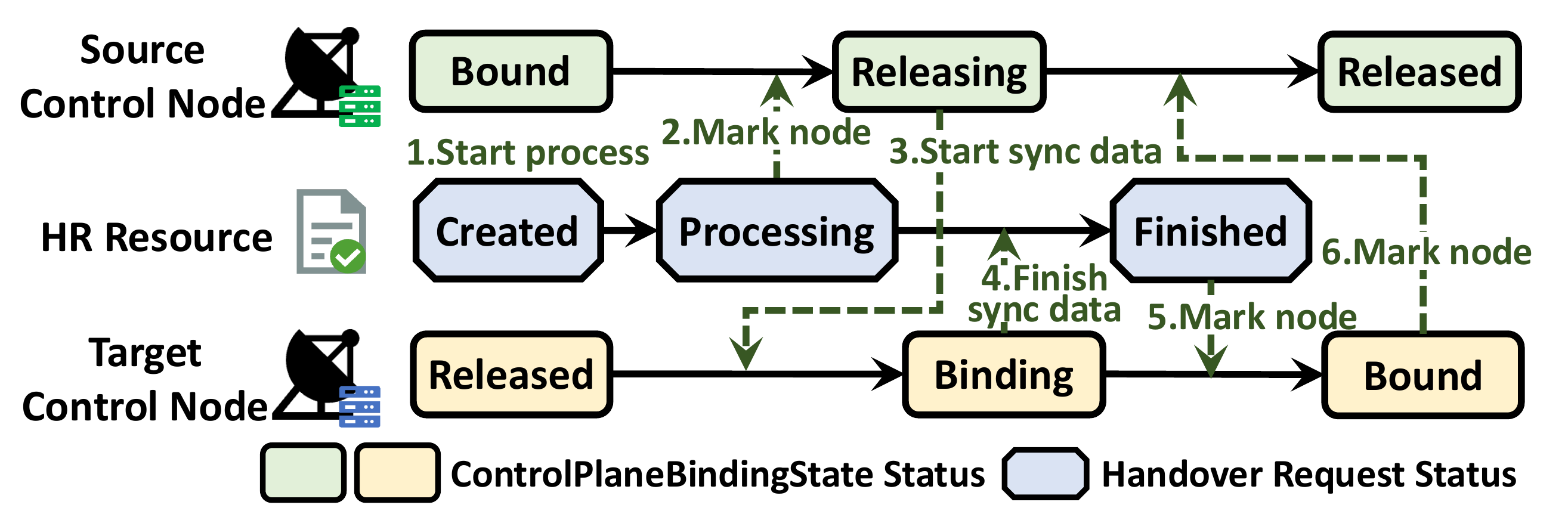}
    \caption{State transition diagram for control node handover.}
    \label{de2}
\end{figure}

\begin{figure}[]
 \vspace{0cm}
	\centering
	\includegraphics[width=1.0\linewidth]{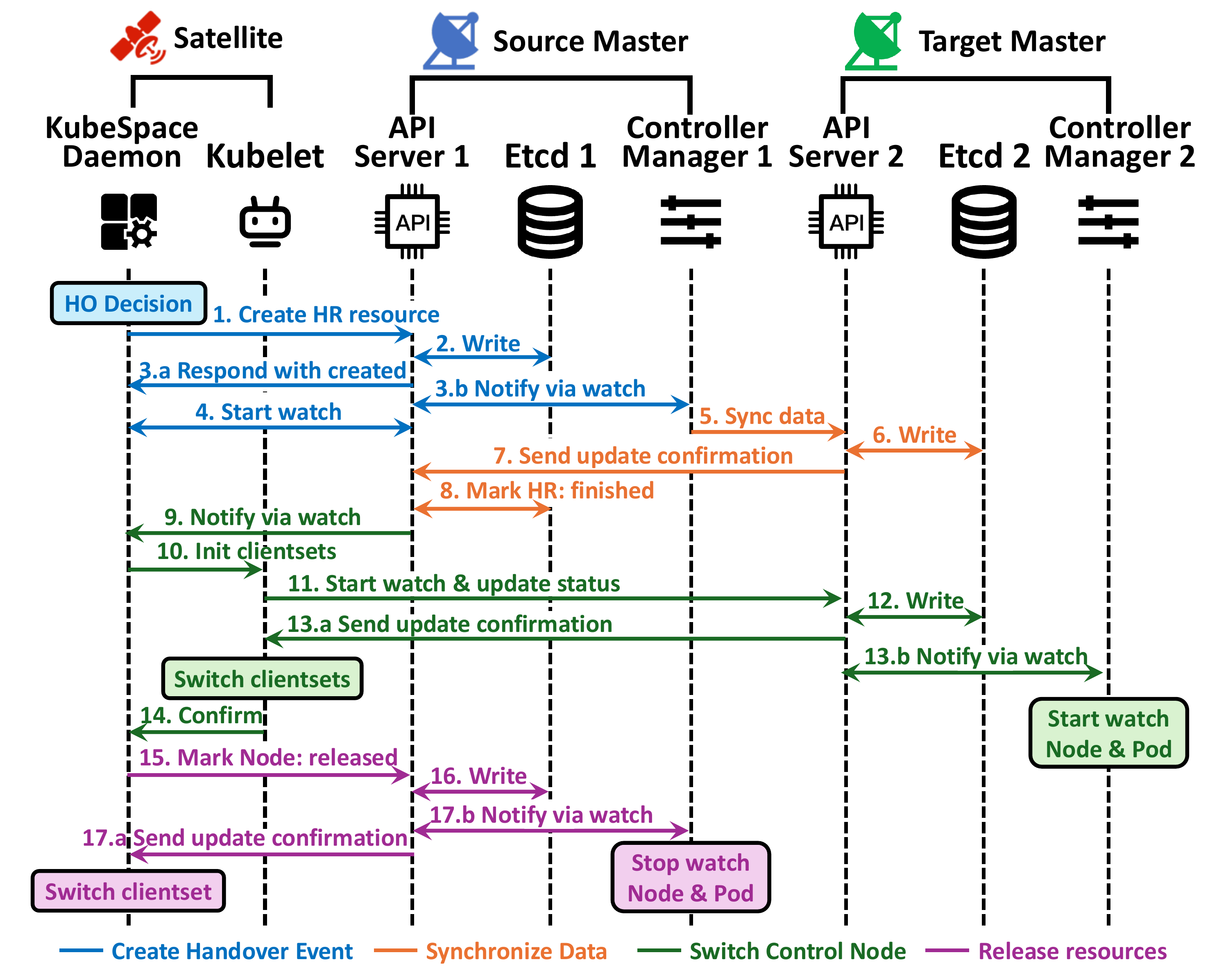}
	\caption{KubeSpace handover procedure.}
	\label{de3}
\end{figure}

Fig.\ref{de2} illustrates the sequence of state transitions in both the HandoverRequest resource and the satellite node’s ControlPlaneBindingState during the handover process. Each state change triggers a corresponding operational step. Once all steps are completed, the ControlPlaneBindingState is updated on both the source and target control nodes, marking the successful transfer of control authority. A complete view of the process is shown in Fig.\ref{de3}, and the detailed procedure is described below.


\noindent\textbf{Steps 1–4}. The KubeSpace daemon on the satellite submits a HandoverRequest resource to the current control node’s API server to initiate the handover process. Once the resource is successfully created, the daemon establishes a watch channel to monitor its status. At the same time, the API server pushes the creation event to the HRController, which continuously watches for resources of this type.

\noindent\textbf{Steps 5–8}. Upon receiving the creation event of the HandoverRequest, the HRController updates the request’s status to Processing and sets the corresponding node’s ControlPlaneBindingState to Releasing. In this state, the node becomes ineligible for new Pod scheduling but continues to report its status to the control node. The HRController then transfers the node and associated Pod configuration data to the target control node’s API server via a synchronous channel. Once the target API server successfully persists the data to etcd, it returns an acknowledgment to the source control node. Finally, the controller manager updates the status of the HandoverRequest to Finished.

\noindent\textbf{Steps 9–14}. The KubeSpace daemon on the satellite detects through the watch channel that the status of the HandoverRequest has changed to Finished. It then instructs the kubelet to load the target control node’s configuration and to initialize a new set of clientsets. The kubelet immediately uses the new clientset to establish a control channel with the target control node and report its current status. Once the target control plane successfully receives this information, it sets the node’s ControlPlaneBindingState to Bound, indicating that control of the node has been formally taken over. After completing the state update, the kubelet switches its internal clientset pointer to the new one and sends a confirmation to the KubeSpace daemon. It then continues to perform state synchronization and task monitoring with the new control node.

\noindent\textbf{Steps 15–17}. The KubeSpace daemon then uses its clientset to update the node’s ControlPlaneBindingState on the source control node to Released. Once this update is successfully committed by the source control node’s API server, KubeSpace switches to the new clientset to communicate with the target control node, thereby marking the formal completion of the handover process.

Throughout the entire control node handover process, the satellite remains visible to the control plane at all times. This is achieved without requiring container migration or Pod eviction. As a result, the services running on the satellite can continue uninterrupted, ensuring availability.

\subsection{Control Node Placement Algorithm}


To systematically analyze the control node placement strategy, we model the problem as an optimization task. Due to the high mobility of LEO satellites, communication latency in satellite-to-satellite and satellite-to-ground links varies dynamically over time. As a result, a sequence of time-varying satellite-ground network topologies emerges over the course of an orbital period. 

Accordingly, we aim to select $K$ ground stations from a set of candidates to serve as control nodes, such that the maximum communication latency between any satellite and its nearest control node across all time instances is minimized. Since propagation delay dominates satellite-ground communication latency, we approximate it using spatial distances between adjacent nodes as edge weights in the network graph to model communication performance.

Let $\tau$ denote the number of considered time instances, and let the set of sampled time points be $T = \{t_1, t_2, \dots, t_\tau\}$. For each time instance $t \in T$, the satellite-ground network topology can be represented as an undirected connected graph $G_t = (V, E_t)$, where the node set is defined as $V = V_s \cup V_g$. Here, $V_s = \{ \text{Sat}_1, \text{Sat}_2, \dots, \text{Sat}_N \}$ denotes the set of satellite nodes, and $V_g = \{ \text{GS}_1, \text{GS}_2, \dots, \text{GS}_M \}$ denotes the set of ground station nodes. Each edge $e_{i,j}^t \in E_t$ indicates a direct link between nodes $i$ and $j$ at time $t$, with the edge weight corresponding to the spatial distance between them. Based on this, we denote $d_{i,j}^t$ as the weighted shortest path length between nodes $i$ and $j$ in graph $G_t$, computed with respect to edge weights.

Suppose we aim to select a set of $K$ control nodes from the $M$ available ground stations. Let $D$ denote the maximum shortest-path distance between any satellite and its nearest control node across all time instances. The definition of $D$ is shown in~(1).
\begin{equation}
D = \max_{t \in T} \max_{s \in V_s} \min_{g \in S} d_{s,g}^t
\end{equation}
where $S \subseteq V_g$ is the selected set of $K$ control nodes. The problem can thus be formulated as a 0--1 integer program that minimizes $D$.

\vspace{-1.1em}

\begin{align}
&\quad \quad \quad \quad \quad \quad \min D \tag{2} \\
\text{s.t.} \quad 
&\quad \quad \quad {\textstyle\sum_{j \in V_g}} d^t_{i,j} \cdot x^t_{i,j} \leq D \tag{3} \\
&\quad \quad \quad {\textstyle\sum_{j \in V_g}} x_{i,j} = 1 \tag{4} \\
&\quad \quad \quad {\textstyle\sum_{j \in V_g}} y_j = K \tag{5} \\
& x_{i,j}^t \leq y_j,\quad x_{i,j}^t \in \left\{ 0,1 \right\} ,\quad y_j \in \left\{ 0,1 \right\} \tag{6}
\end{align}

The model introduces two binary decision variables: $x_{i,j}^t$ and $y_j$. Here, $x_{i,j}^t = 1$ indicates that at time $t$, satellite $i$ selects ground station $j$ as its control node, and $y_j = 1$ indicates that ground station $j$ is selected as a control node. Constraint~(3) defines $D$ as the maximum shortest-path distance between satellites and their corresponding control nodes across all time steps, representing the worst-case communication performance under the current control node placement. Constraint~(4) requires that each satellite be connected to exactly one control node at any given time. Constraint~(5) limits the total number of selected control nodes to $K$.

This problem is essentially an NP-hard 0--1 integer program, and the size of its solution space presents a significant computational challenge. With $M$ candidate ground stations—typically in the hundreds—the number of possible selections of $K$ control nodes is combinatorial, given by $C(M, K) = \binom{M}{K}$. For instance, selecting 10 control nodes from 200 candidates yields $\binom{200}{10} = \frac{200!}{10! \cdot 190!} \approx 2.24 \times 10^{18}$ possible combinations, rendering exhaustive search computationally infeasible. Moreover, capturing the dynamic nature of the satellite--ground topology requires modeling thousands of time steps, during which shortest-path computations between hundreds of ground stations and thousands of satellites must be performed, further compounding the computational complexity.

To obtain a near-optimal solution within a reasonable computational time, we propose a two-stage Control Node Placement Algorithm~(CPNA). First, considering that modern LEO satellite constellations typically exhibit strong structural symmetry and clear orbital periodicity, we introduce an intuitive approach: by clustering satellite–ground network topologies across multiple time instances, we select a small set of representative topologies to serve as proxies for evaluating control node placement. This significantly reduces the computational overhead. Then, based on a greedy strategy, we apply the $k$-center algorithm to select a set of ground stations, followed by a local search procedure to further refine the solution quality. The overall process is summarized in Algorithm~\ref{alg:alg1}.

\begin{algorithm}[t]
\caption{Control Node Placement Algorithm (CNPA)}  \label{alg:alg1} 
\begin{algorithmic}[1]
\Require $\mathcal{G}$: topology snapshots; $k$: number of control nodes; $C$: number of clusters
\Ensure $S$: selected ground stations

\Statex \textbf{Step 1: Topology Clustering}
\State $R \leftarrow$ \textsc{SelectRepresentatives}$(\mathcal{G}, C)$

\Statex \textbf{Step 2: Ground Station Selection}
\State Initialize $S \leftarrow \emptyset$
\For{$i = 1$ to $k$}
    \State $g^* \leftarrow \arg\min_{g \notin S} \textsc{Evaluate}(S \cup \{g\}, \mathcal{G})$
    \State $S \leftarrow S \cup \{g^*\}$
\EndFor
\State $S \leftarrow$ \textsc{LocalSearch}$(S, R)$
\State \Return $S$
\end{algorithmic}
\end{algorithm}

The algorithm consists of three core components, described as follows:

\textsc{SelectRepresentatives}$(\mathcal{G}, C)$:
This function performs clustering on the input set of topology snapshots $\mathcal{G}$. Each topology is first flattened into a feature vector and standardized. Then, $k$-means clustering is applied to partition the topologies into $C$ clusters, aiming to minimize the total intra-cluster variance. From each cluster, the topology closest to the cluster centroid is selected as the representative. The output is a set $R$ of $C$ representative topologies that capture the structural characteristics of $\mathcal{G}$.

\textsc{Evaluate}$(S, \mathcal{G})$:
Given a candidate set of ground stations $S$, the function evaluates the worst-case latency over the entire topology set $\mathcal{G}$. The evaluation metric is defined as $\max_{t \in T'} \max_{s \in V_s} \min_{g \in S} d_{s,g}^t$, which represents the maximum shortest-path distance from any satellite to its nearest ground station in $S$ across all selected time instances. A smaller value implies lower worst-case latency, and thus indicates a more effective control node deployment.

\textsc{LocalSearch}$(S, \mathcal{G})$:
This function iteratively refines the initial candidate set $S$ via local replacements. In each iteration, it attempts to replace one selected ground station in $S$ with an unselected one. If the replacement leads to a lower evaluation score over $\mathcal{G}$, the update is accepted. The process continues until no further improvement is possible or a predefined number of iterations is reached.

\subsection{Control Node Assignment Algorithm}

\begin{algorithm}[t]
\caption{Control Node Assignment Algorithm (CNAA)} \label{alg:alg2}
\begin{algorithmic}[1]
\Require TLE data; $G$: ground control nodes; $T$: time window (e.g., 12 hours); $\Delta_s$: sampling interval (e.g., 60s); $\Delta_p$: decision interval (e.g., 1s); $\delta$: handover threshold ratio
\Ensure $H$: predicted handover events (time, target node)

\Statex \textbf{Step 1: Orbit Prediction and Distance Sampling}
\For{each $t \in [0, T]$ with step $\Delta_s$}
    \State Predict satellite position $p_t$ using TLE
    \For{each $g \in G$}
        \State Compute $d_{t,g} \leftarrow \textsc{Distance}(p_t, g)$
    \EndFor
\EndFor
\State For each $g \in G$, construct interpolation function $f_g(t)$ over $[0, T]$ from $\{d_{t,g}\}$

\Statex \textbf{Step 2: Fine-grained Control Node Assignment}
\State Initialize $H \leftarrow \emptyset$, $g_{\text{curr}} \leftarrow \arg\min_{g} f_g(0)$
\For{each $t \in [0, T]$ with step $\Delta_p$}
    \State $g_{\text{min}} \leftarrow \arg\min_{g} f_g(t)$
    \If{$g_{\text{min}} \neq g_{\text{curr}}$ and $\frac{f_{g_{\text{min}}}(t)}{f_{g_{\text{curr}}}(t)} < \delta$}
        \State $H \leftarrow H \cup \{(t, g_{\text{min}})\}$
        \State $g_{\text{curr}} \leftarrow g_{\text{min}}$
    \EndIf
\EndFor
\State \Return $H$
\end{algorithmic}
\end{algorithm}

After the deployment of ground control nodes is completed, it is necessary to dynamically assign the optimal control node to continuously moving LEO satellites. To this end, this paper designs a Control Node Assignment Algorithm~(CNAA) that can be deployed on satellites. The complete procedure is presented in Algorithm~\ref{alg:alg2}.

This algorithm first introduces an efficient method for estimating the distance between a satellite and ground control nodes over a future time window (e.g., 12 hours). Specifically, it uses the satellite’s TLE orbital data to predict its positions at fixed intervals (e.g., every 60 seconds), and calculates the corresponding distances to all ground control nodes. Based on these sampled values, interpolation functions are constructed to enable efficient and fine-grained satellite-to-ground distance estimation.

Building on this, the algorithm further identifies potential handover events—i.e., the times at which the satellite should handover to a different control node, along with the corresponding target nodes—within the same time window. Handover decisions are primarily based on the estimated spatial distance, with preference given to the closest node at any given time. To prevent excessive handovers caused by minor fluctuations in distance, a threshold mechanism is applied: a handover is triggered only when the candidate node’s distance is shorter than that of the current node by a predefined ratio.

Using spatial distance as the basis for control node selection is straightforward and aligns with the principle of proximity-based management in multi-node systems. Moreover, due to the predictable nature of this assignment mechanism, it can be implemented at the ground station to proactively identify upcoming handover events—such as the satellite involved, the handover time, and the target control node. This enables data pre-synchronization and helps further reduce latency during the handover process.

\section{Performance Evaluation}

In this section, we evaluate whether KubeSpace demonstrates low latency and high stability in managing onboard resources and containers. Additionally, we test the placement and assignment algorithms of control nodes under various satellite constellation scenarios to verify their effectiveness in reducing communication latency and handover frequency between satellites and control nodes.

\subsection{Experimental Setup}

\noindent\textbf{\rev{Constellation and Ground Station Information.}} We simulate three representative LEO satellite constellations based on publicly available data: Starlink~(1,584 satellites)\cite{starlinkLEO}, Kuiper~(1,296 satellites)\cite{kuiperLEO}, and OneWeb~(636 satellites)\cite{onewebLEO}. Each satellite is capable of establishing inter-satellite links with its two adjacent satellites in the same orbit, as well as with the two nearest satellites in neighboring orbits\cite{bhosale2024krios,su2025skyoctopus}. The ground station locations are derived from Starlink’s published deployment data\cite{starlinkGroundstations}, further supplemented by the 200 most populous cities worldwide~\cite{worldUrbanAreas} to increase the flexibility and optionality of deployment strategies.

\noindent\textbf{KubeSpace Prototype.} We built the KubeSpace prototype system using Kubernetes v1.31.10\cite{kubernetes1310} and Go 1.24.3, and set up a simulation platform based on constellation and ground station information for testing purposes. Specifically, we used the Skyfield\cite{skyfield} library's SGP4 orbital model to parse TLE data for various constellations and compute their real-time positions. Satellite and ground control nodes were simulated using KVM virtual machines, which were bound to the simulated ground station and satellite position data. Communication delays were introduced using the tc tool\cite{tc_man8}. The simulation covered the full-day operation of all satellites across three types of constellations. All experiments were conducted on a commercial server (Dell PowerEdge R760) configured with a 3.4 GHz CPU (128 logical cores) and 377 GiB of memory.

\begin{figure*}[t]
\vspace{-0.4cm}
\centering
    \includegraphics[width=0.6\linewidth]{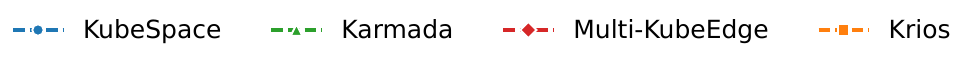}
    \vspace{-0.1cm}
    \subfloat[Starlink]{\includegraphics[width=0.33\textwidth]{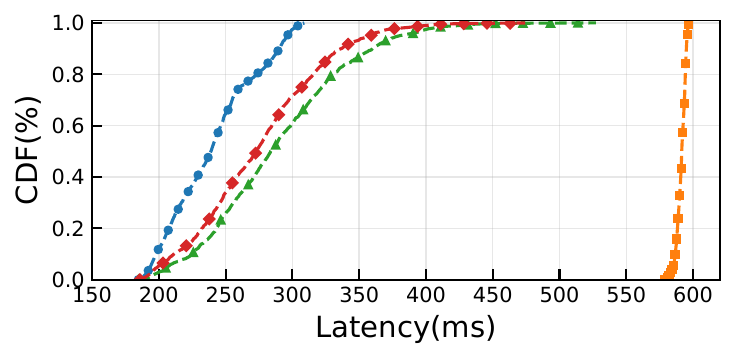}\label{first_starlink}}
    \subfloat[Kuiper]{\includegraphics[width=0.33\textwidth]{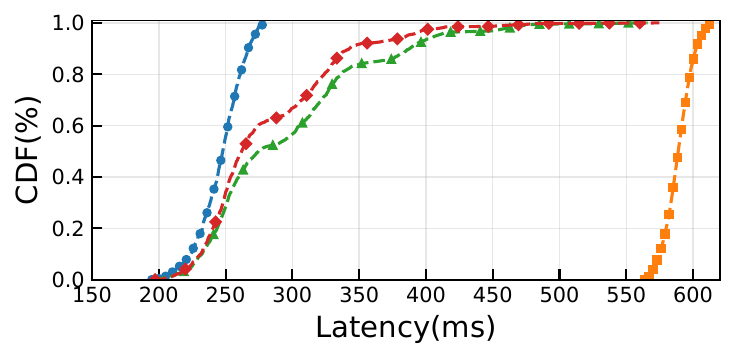}\label{first_kuiper}}
    \subfloat[OneWeb]{\includegraphics[width=0.33\textwidth]{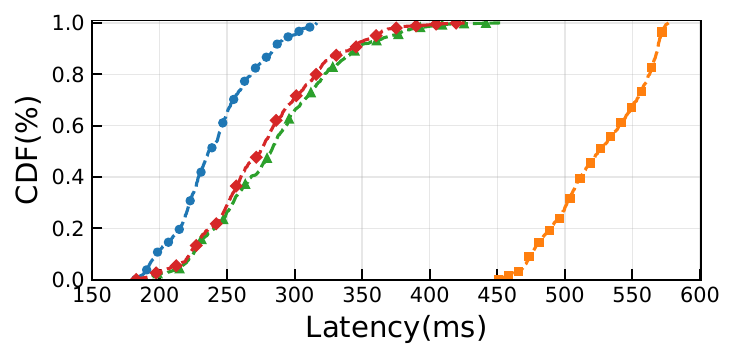}\label{first_oneweb}}
    \vspace{0.cm}
    \caption{Satellite status update latency under different container orchestration platforms.}
    \label{latency_all}
    \vspace{-0.5cm}
\end{figure*}

\noindent\textbf{\rev{Container Orchestration Platforms}.} We evaluated the platform built on the KubeSpace control plane against other leading container orchestration solutions.

\begin{itemize}

\item\textbf{Krios\cite{bhosale2024krios}}: A container orchestration framework for LEO satellites based on a single control node Kubernetes architecture, demonstrating excellent performance in ensuring continuity of user-level container services.

\item\textbf{Karmada\cite{karmada2025}}: A container orchestration platform for managing multiple clusters in ground-based environments. To enable the satellite to restore previously running containers after a control node handover, we manually synchronized satellite data during the transition.

\item\textbf{Multi-KubeEdge}: KubeEdge is a container orchestration platform that has been applied in satellite environments\cite{wang2023satellite2}, offering fault tolerance for disconnected worker nodes. We deployed multiple KubeEdge clusters and manually synchronized satellite data between control nodes during control node handovers, thereby restoring normal operation of onboard containers.

\end{itemize}

\noindent\textbf{\rev{Control Node Placement Algorithm.}} \rev{We compare the control node placement algorithm proposed in KubeSpace with the following two approaches to evaluate its effectiveness.}

\begin{itemize}

\item \textbf{Exhaustive Optimal}: We traverse all possible combinations of the K ground stations and select the one that minimizes the maximum satellite-to-control-node latency across all time steps.
\item \textbf{Random Selection}: We randomly select K ground stations from all candidates as control nodes, serving as a baseline for the multi control node placement scheme.

In addition, we select a city near the equator as the baseline for the single control node placement.

\end{itemize}

\noindent\textbf{\rev{Control Node Assignment Algorithm}.} We compare the control node assignment algorithm proposed by KubeSpace with a baseline that selects the control node solely based on the shortest distance, and analyze the differences between the two in terms of communication latency and handover frequency.

\begin{figure}[t]
 \vspace{0cm}
    \centering
    \includegraphics[width=1.0\linewidth]{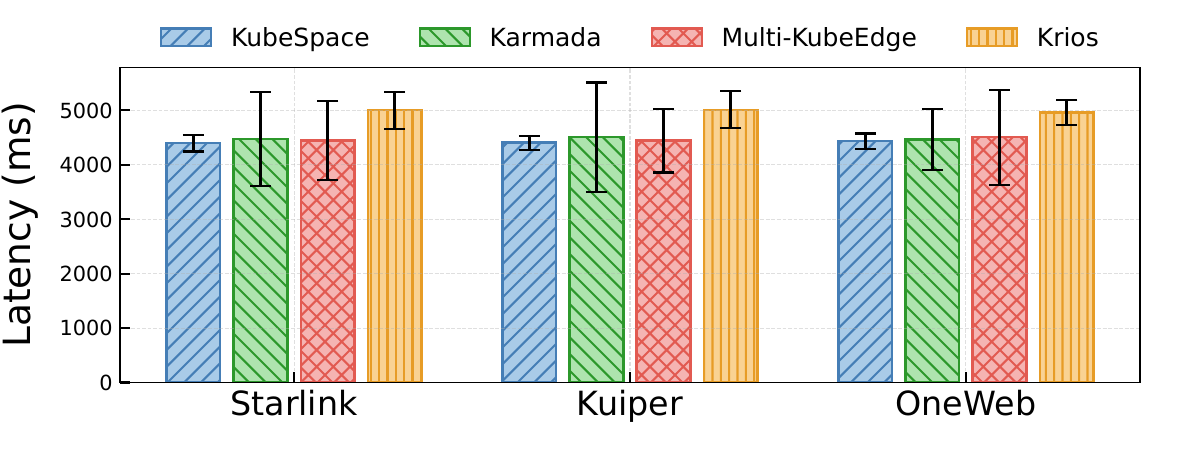}
    \vspace{-0.7cm}
    \caption{Pod recovery time under different container orchestration platforms..}
    \label{event_response_time}
\vspace{-0.15cm}

\end{figure}

\subsection{Experimental Results}

We first compared the container orchestration platform based on the KubeSpace control plane with three other container orchestration platforms. For Karmada and Multi-KubeEdge, which also adopt a multi control node architecture, we simulated the same 10 ground station locations for deploying control nodes. For Krios, which uses a single control node, we placed its ground station in a city near the equator to minimize the maximum communication latency with satellites. In terms of satellite node management, we conducted comparisons from two aspects: node status reporting and in-orbit node event response. Additionally, for the multi control node solutions, we evaluated the performance overhead introduced by control node handovers.


\noindent\textbf{Satellite Status Reporting Latency.} Timely status reporting is a key factor in ensuring the stable operation of container orchestration platforms. Worker nodes report the status of nodes and containers to the control plane either periodically or when triggered by specific events. We measured the reporting latency and calculated the average status reporting latency for each satellite over a single operational cycle, and further plotted the corresponding cumulative distribution function. 

As shown in Fig.~\ref{latency_all}, the node status reporting latency in KubeSpace is significantly lower than that in Krios. Taking Starlink as an example, the average value of this metric across all satellites in KubeSpace is 240ms, compared to 591ms in Krios—a reduction of approximately 59\%. For Karmada and Multi-KubeEdge, some of the selected reporting timestamps fall within the control node handover period, requiring satellites to wait for the handover to complete before completing the status report. This results in higher reporting latency compared to KubeSpace. In terms of the maximum average status reporting latency, KubeSpace achieves reductions of 41\% and 34\% compared to Karmada and Multi-KubeEdge, respectively.

\noindent\textbf{Response Time to Node Pressure.} An essential capability of container orchestration platforms is the prompt migration of containers when a node experiences resource pressure, thereby preventing service disruption. We evaluated the response time of various platforms under such scenarios. Specifically, we preloaded the YOLOv5~\cite{shangguang2024first} container image onto all satellite nodes and initially launched the container on one of them. At a randomly chosen time, we triggered a resource pressure event on the active satellite, thereby initiating the container migration to a neighboring satellite. We define the response time as the duration between the triggering of the event and the resumption of the container on the neighboring satellite.

As shown in Fig.~\ref{event_response_time}, KubeSpace achieves the shortest average container recovery time, reducing it by approximately 12\% compared to Krios. For Karmada and Multi-KubeEdge, since the time point at which we triggered node resource contention partially overlapped with the control plane handover window, they had to wait for the handover process to complete before proceeding with container scheduling. This resulted in a higher standard deviation of recovery time compared to KubeSpace. Moreover, considering that container startup itself takes around 4 seconds, the advantage of a multi control node architecture is less pronounced in this experiment than in the status reporting experiment.

\begin{figure}[t]
  \centering

  \subfloat[Handover duration.]{%
    \includegraphics[width=0.47\linewidth]{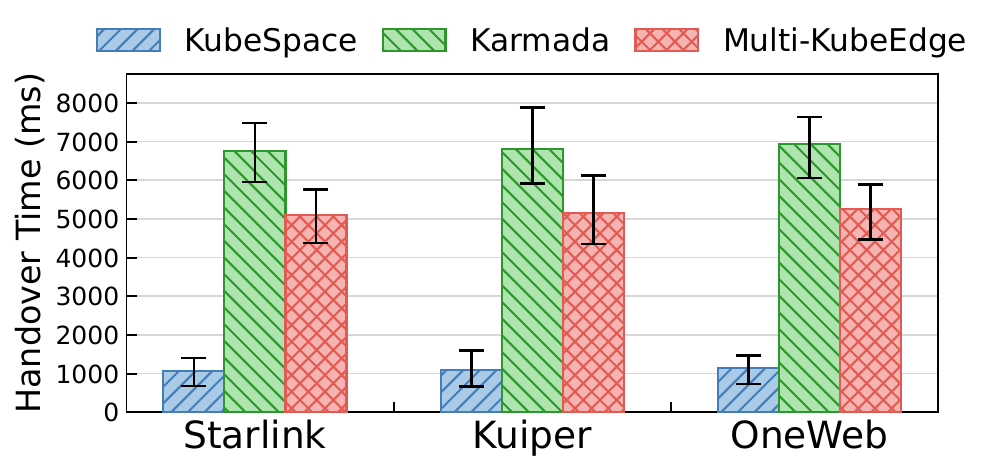}%
    \label{metric1}
  }
  \hspace{0.01\linewidth}
  \subfloat[Daily cumulative handover time.]{%
    \includegraphics[width=0.47\linewidth]{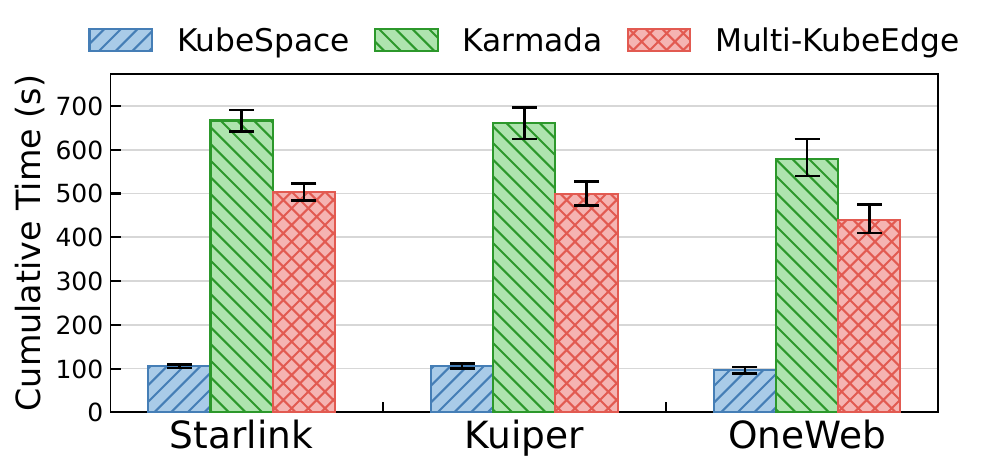}%
    \label{metric2}
  }\\[5pt]

  \subfloat[Daily cumulative node invisibility time.]{%
    \includegraphics[width=0.47\linewidth]{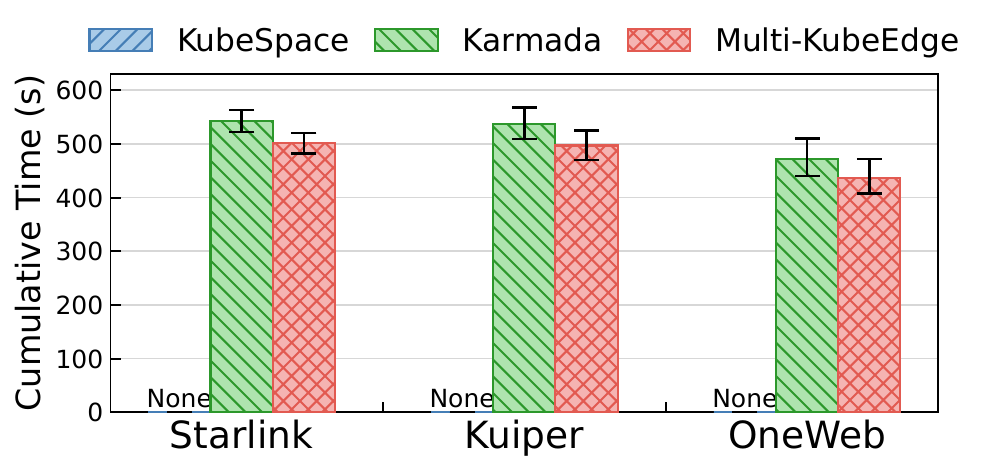}%
    \label{metric3}
  }
  \hspace{0.01\linewidth}
  \subfloat[Daily cumulative pod unavailability time.]{%
    \includegraphics[width=0.47\linewidth]{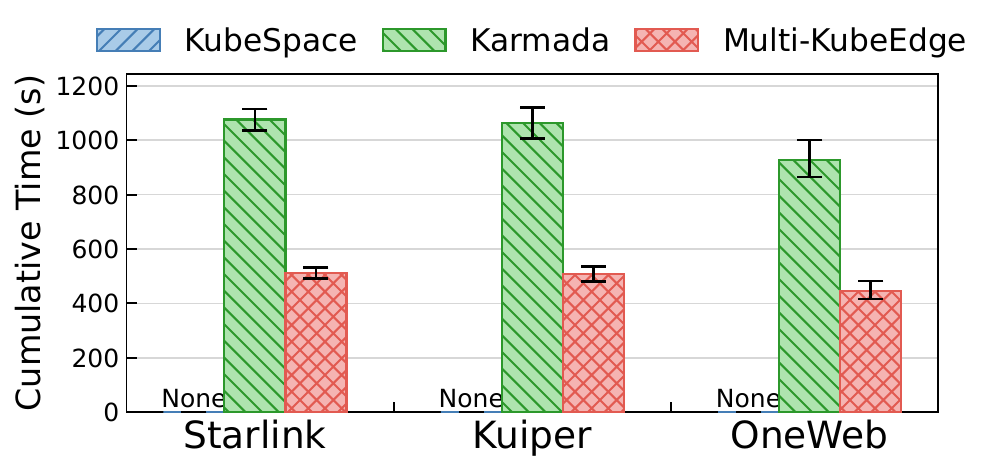}%
    \label{metric4}
  }

  \caption{Overhead of control node handovers (per satellite).}
  \label{fig:combined_metrics}
\vspace{-0.15cm}
\end{figure}

\noindent\textbf{Handover Time Overhead.} We conducted a comparative analysis of the handover latency across three multi control node container orchestration platforms. As shown in Fig.\ref{metric1} and Fig.\ref{metric2}, KubeSpace outperforms the other two solutions significantly in terms of both single handover latency and the cumulative handover latency per satellite over a 24-hour period. Compared to Karmada and Multi-KubeEdge, KubeSpace reduces handover latency by approximately 84\% and 79\%, respectively. Multi-KubeEdge achieves slightly lower latency than Karmada, owing to its support for local offline caching of Pod configurations, which eliminates the need for Pod eviction during cluster exit and remote fetching of previously running Pods upon rejoining the cluster.

\noindent\textbf{Satellite Invisibility Duration.} Fig.~\ref{metric3} compares the duration of satellite node invisibility caused by control plane handovers. With KubeSpace, the Kubelet continuously reports node status to at least one control node during the handover, allowing \texttt{kubectl get node} to consistently retrieve node information. As a result, the node invisibility duration is effectively reduced to zero. In contrast, both Karmada and Multi-KubeEdge experience a temporary disconnection and reconnection process during control plane handovers, leading to an average daily invisibility duration of 543 and 502 seconds per satellite, respectively.

\noindent\textbf{Pod Unavailability Duration.} We deployed a static content service based on nginx on the satellites and continuously tested its accessibility via GET requests\cite{bhosale2024krios}. As shown in Fig.~\ref{metric4}, since KubeSpace does not restart containers during the handover process, the container service remains continuously available. In contrast, under the Karmada and Multi-KubeEdge schemes, the average cumulative daily service downtime per satellite reaches 1076 seconds and 512 seconds, respectively.

\begin{figure}[t]
\vspace{-0.3cm}
  \centering
  \includegraphics[width=\linewidth]{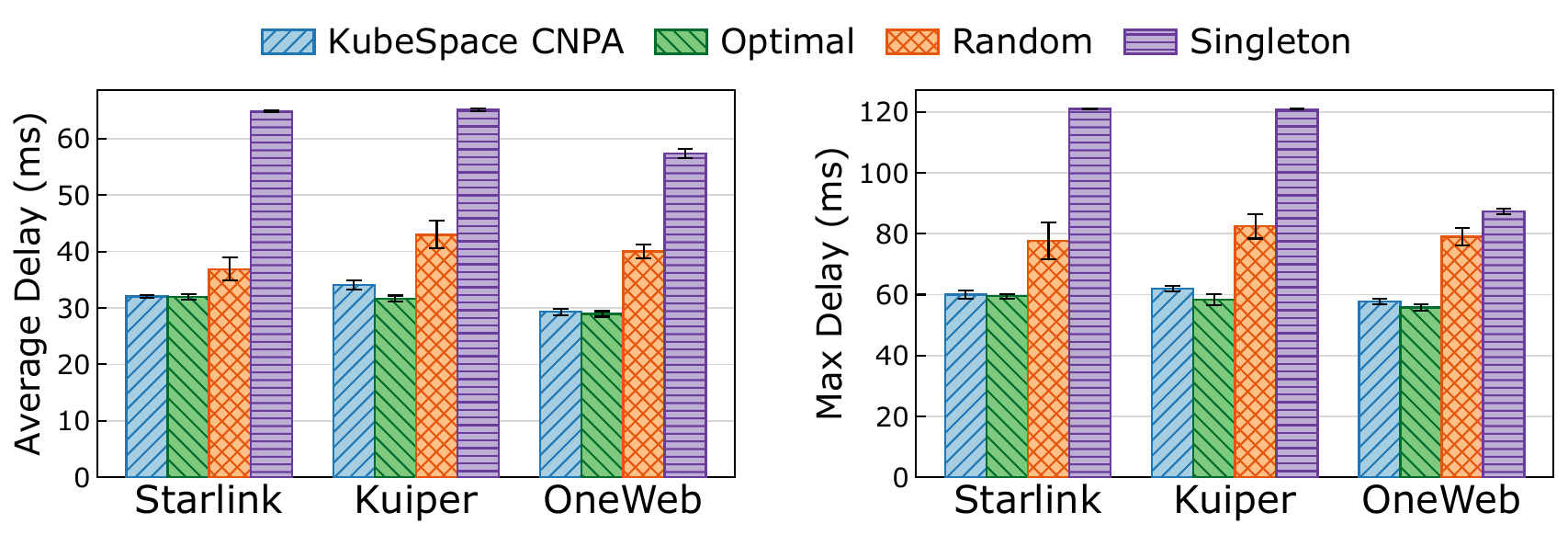}
  \caption{Comparison of control node placement methods.}
  \label{data1}
\vspace{-0.15cm}
\end{figure}

\noindent\textbf{Satellite-to-Control-Node Communication Latency.} As shown in Fig.~\ref{data1}, with $K = 5$ and 20 topology clusters, we compare the CNPA strategy provided by KubeSpace with three other methods in terms of the average and maximum communication latency between all satellites and the control node over the course of a day. Taking the Starlink constellation as an example, the strategy generated by CNPA results in an average daily maximum communication latency of 60.05\,ms, which is within 1\,ms of the optimal value obtained via exhaustive search. In comparison, this value is reduced by 22\% and 50\% compared to the random scheme and the single-node scheme, respectively. The average communication latency is 32.05\,ms, also within 1\,ms of the optimal value.

As $k$ increases, CNPA still yields high-quality solutions within reasonable time despite the rising cost of exhaustive search, effectively reducing average communication latency. Next, we evaluate the assignment algorithms under varying numbers of control nodes (2 to 20), based on CNPA’s placement schemes.

\begin{figure}[t]
  \vspace{-0.3cm}
  \centering
  \includegraphics[width=0.98\linewidth]{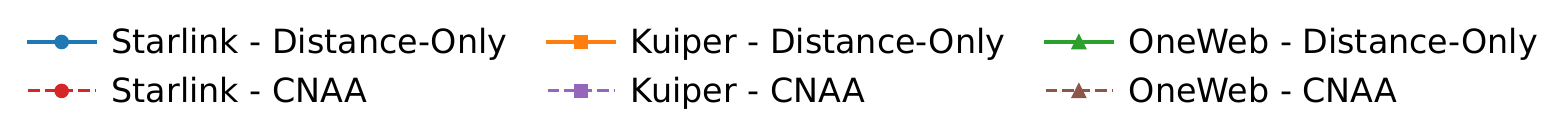}
  \vspace{-0.1cm}
  \subfloat[Daily handovers per sat.]{%
    \includegraphics[width=0.48\linewidth]{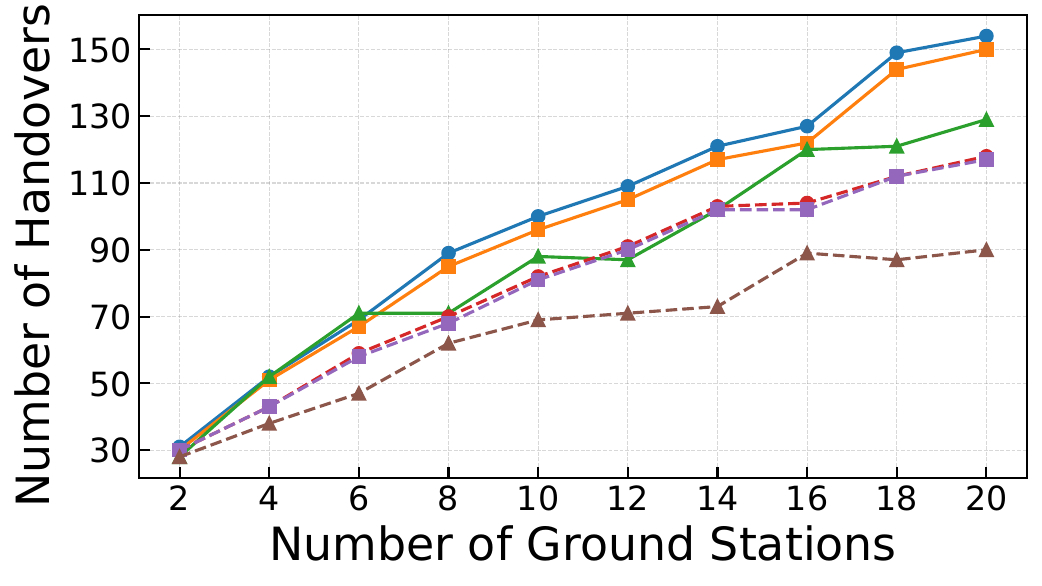}
    \label{cnaa_data1}
  }
  \subfloat[Avg. latency: sat–control node.]{%
    \includegraphics[width=0.48\linewidth]{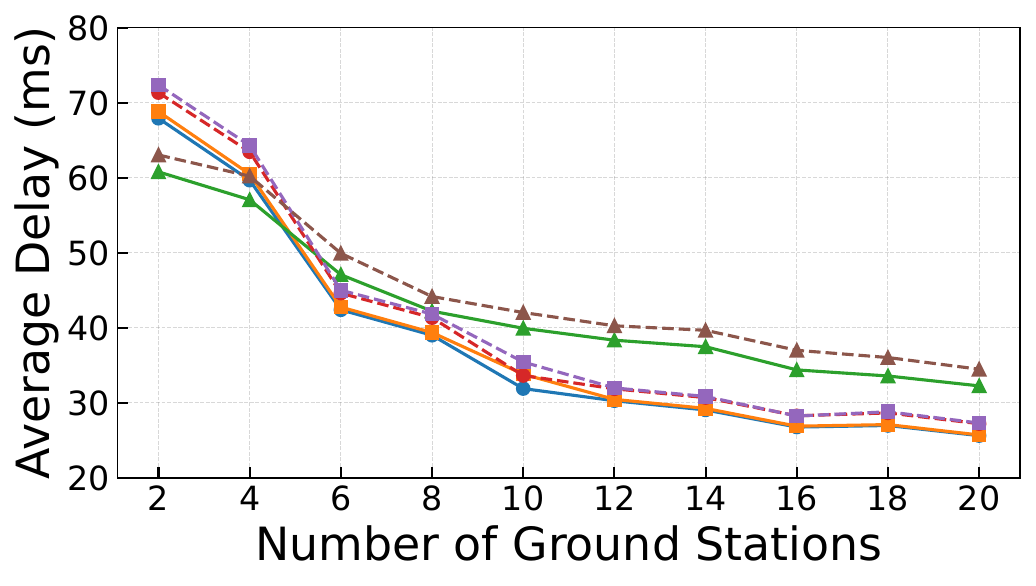}
    \label{cnaa_data2}
  }
  \caption{Average latency and handover frequency under different control node assignment methods.}
  \label{cnaa}
\vspace{-0.15cm}
\end{figure}


\noindent\textbf{\rev{Relationship Between \needrev{Handover} Frequency and Latency of CNNA.}}  Fig.~\ref{cnaa} \rev{plots evaluation of the proposed Control Node Assignment Algorithm in KubeSpace under varying numbers of ground stations. For the Starlink constellation, CNAA achieves a 19\% reduction in control‑node handover frequency while incurring only an additional 2~ms in average communication latency. These results demonstrate that CNAA attains a superior trade‑off between communication latency and handover overhead compared with the shortest‑distance‑only baseline.}

\section{Related Work}

In recent years, researchers have explored computing paradigms like container orchestration in LEO satellites\cite{wang2023satellite1,shangguang2024first,wang2023satellite2,bhosale2024krios,pfandzelter2024komet,pusztai2024hyperdrive,yan2019satec,pfandzelter2021towards}. For example, \cite{shangguang2024first,wang2023satellite2} deployed AI applications using KubeEdge \cite{kubeedge} on satellites, validating the orchestration feasibility in space. \cite{bhosale2024krios} proposed a scheduling mechanism for LEO, and \cite{pfandzelter2024komet} designed a stateless edge platform to improve onboard service availability. While these studies optimize for LEO characteristics, they often overlook the challenge of achieving both low latency and high stability in the control plane. This work addresses that gap with a proposed solution.

With the growing adoption of the space cloud computing paradigm, researchers have increasingly focused on improving onboard computing resource utilization through effective task scheduling strategies\cite{liu2024orbit,denby2023kodan,tong2022joint,ji2023cooperative,wang2023energy}. \cite{liu2024orbit} proposed a solar-illumination-aware scheduling algorithm to reduce energy consumption and extend satellite lifespan, while \cite{denby2023kodan} introduced a task offloading mechanism to enhance performance and meet timeliness requirements. Although these studies made notable progress in scheduling strategies, they generally overlook the practical challenges of deploying and maintaining the control plane in space. To address this, KubeSpace builds a low-latency, high-stability control plane that supports real-world deployment of such strategies onboard satellites.

Software-Defined Satellite Networks (SDRN) aim to enhance flexibility and programmability by decoupling control and data planes\cite{jiang2023software,bao2014opensan,zhang2021space}, but existing work largely remains theoretical with limited focus on practical deployment. In contrast, KubeSpace provides a deployable and stable control plane for efficient onboard computing resource management.

\section{Discussion and Limitation}




Although this paper presents a container orchestration control plane for LEO satellites that balances low latency and stability, building an efficient, resilient space cloud platform still poses many challenges requiring further investigation. For example, in terrestrial orchestration systems, replica management is key to service availability. Prior work (e.g., \cite{bhosale2024krios}) has explored improving availability via service migration, but how to efficiently leverage replicas for redundancy in orbit remains open. Moreover, pulling and distributing container images can become a deployment bottleneck due to limited satellite-ground bandwidth. Optimizing image transmission through caching, delta synchronization, and related techniques is another important area for future research.

\section{Conclusion}

Existing container orchestration platforms face challenges in meeting both low latency and high stability requirements when applied to LEO satellite constellations characterized by wide spatial distribution and frequent handovers. To address this, we propose KubeSpace. This approach leverages an optimized multi control node architecture and an efficient control node deployment and allocation mechanism, reducing the latency of onboard container management by 59\% compared to traditional solutions and effectively eliminating node management interruptions present in previous approaches. As a potential future direction, we are looking forward to extending our method to improve the performance of various applications such as large language models~\cite{fang2024automated,lin2023pushing,qu2025mobile} and distributed learning system~\cite{lin2024adaptsfl,zhang2025lcfed,wei2025pipelining,hu2024accelerating,lin2024efficient,zhang2024fedac,zhan2025prism}.

\bibliographystyle{IEEEtran}
\bibliography{ref.bib}

\end{document}